\documentclass[11pt]{article}
\usepackage{osid}
\usepackage{eepic,psfig,pstricks}
\usepackage{fancybox}
\usepackage[dvips]{graphicx}
\usepackage{amsmath,amssymb,mathrsfs}
\usepackage{epic,eepic}
\usepackage{epsfig}
\def\Tr{\operatorname{Tr}} 
\def\>{\rangle}\def\<{\langle} 
   
\def\openone{I}
\begin{document}

\title{A minimum-disturbing quantum state discriminator}

\author{Francesco Buscemi
\\{\footnotesize\it Daini Hongo White Blgd. 201, 5-28-3 Hongo, Bunkyo-ku, 113-0033 Tokyo, Japan \& buscemi@qci.jst.go.jp}\\[2ex]
  Massimiliano F. Sacchi \\{\footnotesize\it
    Dip. di Fisica ``A. Volta'' and CNISM, via Bassi 6, I-27100 Pavia,
    Italy \& msacchi@unipv.it}}

\maketitle
\begin{abstract}
We propose two experimental schemes for quantum state discrimination
that achieve the optimal tradeoff between the probability of correct
identification and the disturbance on the quantum state.
\end{abstract}

%%%%%%%%%%%%%%%%%%%%%%%%%%%%%%%%%%%%%%%%%%%%%%%%%%%%%%%%%%%%%%%%%%%%%%
\section{Introduction}
Indistinguishability of nonorthogonal states is a basic feature of
quantum mechanics that has deep implications in many areas, as quantum
computation and communication, quantum entanglement, cloning, and
cryptography. Since the pioneering work of Helstrom \cite{helstrom} on
quantum hypothesis testing, the problem of discriminating
nonorthogonal quantum states has received a lot of attention
\cite{rev12}, with some experimental verifications as well
\cite{exper}. The most popular scenarios are: $i)$ the minimum-error
probability discrimination \cite{helstrom}, where each measurement
outcome selects one of the possible states and the error probability
is minimized; $ii)$ the optimal unambiguous discrimination
\cite{unam}, where unambiguity is paid by the possibility of getting
inconclusive results from the measurement; $iii)$ the minimax strategy
\cite{mmax} where the smallest of the probabilities of correct
detection is maximized. Stimulated by the rapid developments in
quantum information theory, the problem of discrimination has been
addressed also for bipartite quantum states, along with the comparison
of global strategies where unlimited kind of measurements is
considered, with the scenario of LOCC scheme, where only local
measurements and classical communication are allowed \cite{walg}.  The
concepts of nonorthogonality and distinguishability can be applied
also to quantum operations, namely all physically allowed
transformations of quantum states, and some work has been devoted to
the problem of discriminating unitary transformations \cite{CPR} and
more general quantum operations \cite{discrcp}.

The quantum indistinguishability principle is closely related to
another very popular, yet often misunderstood, principle (formerly
known as Heisenberg principle~\cite{heisenberg,fuco,banaszek01.prl}):
it is not possible to extract information from a quantum system
without perturbing it somehow. In fact, if the experimenter could
gather information about an unknown quantum state without disturbing
it at all, even if such information is partial, by performing further
non-disturbing measurements on the same system, he could finally
determine the state, in contradiction with the indistinguishability
principle~\cite{gdm-yuen}.

Actually, there exists a precise tradeoff between the amount of
information extracted from a quantum measurement and the amount of
disturbance caused on the system, analogous to Heisenberg relations
holding in the preparation procedure of a quantum state.  Quantitative
derivations of such a tradeoff have been obtained in the scenario of
quantum state estimation \cite{hol,qse}. The optimal tradeoff has been
derived in the following cases: in estimating a single copy of an
unknown pure state \cite{banaszek01.prl}, many copies of identically
prepared pure qubits \cite{banaszek01.pra}, a single copy of a pure
state generated by independent phase-shifts \cite{mista05.pra}, an
unknown maximally entangled state \cite{max}, an unknown coherent
state \cite{cv} and Gaussian state \cite{paris}, and an unknown spin
coherent state \cite{maxun}.
 
Experiment realization of minimal disturbance measurements has been
also reported \cite{cv,dema}.

In the present paper we review the characterization of the tradeoff
relation in quantum state discrimination of Ref. \cite{qph}, and
suggest an experimental realization of the minimum-disturbing
measurement.  In this case, an unknown quantum state is chosen with
equal \emph{a priori} probability from a set of two non orthogonal
pure states, and the error probability of the discrimination is
allowed to be suboptimal (thus intuitively causing less disturbance
with respect to the optimal discrimination). A measuring strategy that
achieves the optimal tradeoff is shown to smoothly interpolate between
the two limiting cases of maximal information extraction and no
measurement at all.  The issue of the information-disturbance tradeoff
for state discrimination can become of practical relevance for posing
general limits in information eavesdropping and for analyzing security
of quantum cryptographic communications. 

After briefly reviewing the optimal information-disturbance tradeoff
in quantum state discrimination and the corresponding measurement
instrument, we analyze two possible experimental realization of the
minimum-disturbing measurement.

\section{Information-disturbance tradeoff in quantum state discrimination}

Typically, in quantum state discrimination we are given two (fixed)
non orthogonal pure states $\psi_1$ and $\psi_2$, with \emph{a priori}
probabilities $p_1$ and $p_2=1-p_1$, and we want to construct a
measurement discriminating between the two. We can describe a measurement 
by means of an \emph{instrument} \cite{davies-lewis}, namely,
a collection of completely positive maps $\{\mathcal{E}_i\}$, labelled
by the measurement outcomes $\{i\}$. Using 
the Kraus decomposition~\cite{kraus}, one can always write
$\mathcal{E}_i(\rho)=\sum_kE^{(i)}_k\rho E^{(i)\dag}_k$. In the case
the sum comprises just one term, namely, $\mathcal{E}_i(\rho)=E_i\rho
E^\dag_i$, the map $\mathcal{E}_i$ is called \emph{pure}, since it
maps pure states into pure states. The trace
$\Tr[\mathcal{E}_i(\rho)]=\Tr[\Pi_i\rho]$, where
$\Pi_i=\sum_kE^{(i)\dag}_kE^{(i)}_k$ is a positive operator associated
to the $i$-th outcome, provides the probability that the measurement
performed on a quantum system described by the density matrix $\rho$
gives the $i$-th outcome. The posterior (or reduced) state after the
measurement is given by
$\rho_i=\mathcal{E}_i(\rho)/\Tr[\mathcal{E}_i(\rho)]$. The averaged
reduced state---coming from ignoring the measurement outcome---is
simply obtained using the \emph{trace-preserving} map
$\mathcal{E}=\sum_i\mathcal{E}_i$. The trace-preservation constraint
for $\mathcal{E}$ implies that the set of positive operators
$\{\Pi_i\}$ is actually a positive operator-valued measure (POVM),
satisfying the completeness condition $\sum_i\Pi_i=\openone$.

Quantum state discrimination is then performed by a two-outcome
instrument $\{\mathcal{E}_1,\mathcal{E}_2\}$ whose capability of
discriminating between $\psi_1$ and $\psi_2$ can be evaluated by the
average success probability
\begin{equation}\label{eq:prob}
  P(\{\mathcal{E}_1,\mathcal{E}_2\})
=\sum_{i=1}^2p_i
  \Tr[\mathcal{E}_i(|\psi_i\>\<\psi_i|)]
=\sum_{i=1}^2p_i\Tr[\Pi_i|\psi_i\>\<\psi_i|].
\end{equation}
Notice that $P$ actually depends only on the POVM $\{\Pi_i\}$. The
probability $P$ quantifies the amount of information that the
instrument $\{\mathcal{E}_1,\mathcal{E}_2\}$ is able to extract from
the ensemble $\{p_1,\psi_1;p_2,\psi_2\}$. Among all instruments
achieving average success probability $\bar P$ (the bar over $P$ means
that we fix the value of $P$), we are interested in those minimizing
the average disturbance caused on the unknown state, that we evaluate
in terms of the average fidelity, namely,
\begin{equation}\label{eq:disturb}
  D(\{\mathcal{E}_1,\mathcal{E}_2\},\bar P)=1-\sum_{i=1}^2p_i\<\psi_i|\mathcal{E}(|\psi_i\>\<\psi_i|)|\psi_i\>.
\end{equation}
Differently from $P$, the disturbance $D$ strongly depends on the
particular form of the instrument $\{\mathcal{E}_i\}$. This means that
there exist many different instruments achieving the same $P$, but
giving different values of $D$. Let
\begin{equation}\label{eq:hard-min}
  \bar D(\bar
  P)=\min_{\{\mathcal{E}_1,\mathcal{E}_2\}}
  D(\{\mathcal{E}_1,\mathcal{E}_2\},\bar P) \end{equation} be the
  disturbance produced by the \emph{least disturbing} instrument that
  discriminates $\psi_1$ from $\psi_2$ with average success
  probability $\bar P$. Intuitive arguments suggest that the larger is
  $\bar P$, the larger must correspondingly be $\bar D$ (i.~e., the
  larger is the amount of information extracted, the larger is the
  disturbance caused by the measurement). The precise derivation of
  the optimal tradeoff $\bar D(\bar P)$ has been obtained in
  Ref. \cite{qph}, along with the corresponding optimal
  measurement, for 
equal \emph{a priori} probabilities, i.~e. $p_1=p_2=1/2$. 
In the following we briefly review the main results. 

\begin{figure}[htb]
\begin{center}
\epsfig{file=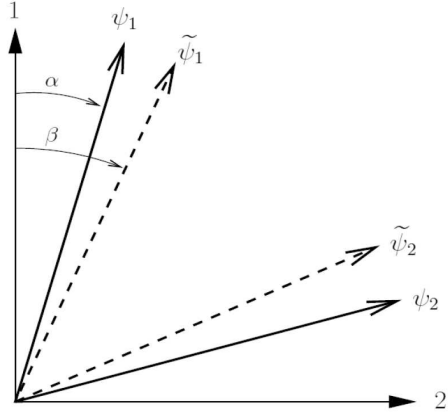,width=6cm}
\caption{Helstrom's scheme to optimally discriminate between to non
  orthogonal states $\psi_1$ and $\psi_2$. The orthogonal axes $1$ and
  $2$ correspond to the von Neumann measurement that 
  achieves the optimal discrimination probability
  (\ref{eq:helstrom}). According to the measurement outcome, 
  $\widetilde\psi_1$ and $\widetilde\psi_2$ are the states to
  be prepared, in order to
  minimize the disturbance.}
\label{fig:quadrante}
\end{center}
\end{figure}

Let us start reviewing the case of the measurement maximizing
$P$. Notice that, given two generally non orthogonal pure states
$\psi_1$ and $\psi_2$, it is always possible to choose an 
orthonormal basis $\{|1\>,|2\>\}$, placed symmetrically around
$\psi_1$ and $\psi_2$ (see Fig.~\ref{fig:quadrante}), on which both
states have real components, namely
\begin{equation}
\begin{split}\label{eq:input-states}
  &|\psi_1\>=\cos\alpha\;|1\>+\sin\alpha\;|2\>,\\
  &|\psi_2\>=\sin\alpha\;|1\>+\cos\alpha\;|2\>,
\end{split}
\end{equation}
and fidelity $f =|\<\psi_1|\psi_2\>|=\sin 2\alpha$. In this case,
it is known~\cite{helstrom} that the maximum achievable $P$ is given by 
\begin{equation}\label{eq:helstrom}
  P_\textrm{opt}=\cos^2\alpha  ,
\end{equation}
which is obtained by the orthogonal von Neumann measurement
$\{|1\>\<1|,|2\>\<2|\}$.

The instrument achieving
$P_\textrm{opt}$, that minimizes the disturbance $D$ is given by 
\begin{equation}\label{eq:Ui}
  \mathcal{E}_i(\rho)=U_i|i\>\<i|\rho|i\>\<i|U^\dag_i,\qquad i=1,2,
\end{equation}
where $U_1$ is the unitary operator
\begin{equation}
U_1=\begin{pmatrix}
\cos\beta & \sin\beta\\
-\sin\beta & \cos\beta
\end{pmatrix},\label{u1}
\end{equation}
$U_2=U_1^\dag $, and $\beta $ satisfies the equation \cite{fuco,qph} 
\begin{equation}\label{eq:tilt}
\tan2\beta=\frac{\tan2\alpha}{\cos2\alpha}.
\end{equation}
Equation (\ref{eq:Ui}) represents a 
measure-and-prepare realization: the observable $|i\>\<i|$ is measured
and, depending on the outcome, the quantum state 
$|\widetilde \psi _i \rangle =U_i|i\>$ is
prepared. The states $|\widetilde \psi _i \rangle $ are 
symmetrically tilted with respect to the $|\psi_i \rangle $'s, see
Fig.~\ref{fig:quadrante}.  

The presence of the tilt $\beta$ can be understood by noticing that
minimum error discrimination can never be error-free for non
orthogonal states. Even using the optimal Helstrom's
measurement, there is always a non zero error probability, and, the
closer the input states are to each other, the smaller the success
probability is. Hence,  it is reasonable that, the closer the input
states are, the less ``trustworthy'' the measurement outcome is, and
the average disturbance is minimized by cautiously preparing a new
state that actually is a coherent superpositions of both hypotheses
$\psi_1$ and $\psi_2$. The minimum disturbance for
Helstrom's optimal measurement is given by
\begin{equation}\label{eq:disturbance_a}
  D_\textrm{opt}=\frac{4-\sqrt{14+2\cos 8\alpha }}{8}.
\end{equation}
Notice that $D_\textrm{opt}$ reaches its maximum for $\alpha=\pi/8$,
namely, when $\psi_1$ and $\psi_2$ are ``unbiased'' with respect to
each other ($|\<\psi_1|\psi_2\>|^2=1/2$).

By allowing a suboptmal discrimination, with success 
probability $P < P_\textrm{opt}$, one can cause less disturbance. 
In this case, by parametrizing the average success
probability thorugh a control parameter $t$ as follows
\begin{equation}\label{eq:prob1-t}
  P_t=tP_\textrm{opt}+\frac{1-t}{2} 
= t\cos^2 \alpha +\frac{1-t}{2},
\end{equation}
with $0\le t\le 1$, one can search, among all possible
measurements achieving $P_t$, the one minimizing the disturbance
$D(P_t)$. It turns out that, for any value of  
$P_t$, the minimum disturbance $D_t$ is achieved by the \emph{pure}
instrument $\mathcal{E}_i^{(t)}(\rho)=E_i^{(t)}\rho E_i^{(t)\dag }$,
where \cite{qph}
\begin{equation}\label{eq:kraus-t}
\begin{split}
  E_1^{(t)}&=U(t)\left(\frac{\sqrt{1-\gamma}}{2}\sigma_z+\frac{\sqrt{1+\gamma}}{2}\openone\right),\\
  E_2^{(t)}&=U^\dag(t)\left(-\frac{\sqrt{1-\gamma}}{2}\sigma_z+\frac{\sqrt{1+\gamma}}{2}\openone\right),
\end{split}
\end{equation}
with $\gamma=\sqrt{1-t^2}$.  The unitary operator $U(t)$ in the above
equation generalizes that in Eq.~(\ref{u1}) as follows
\begin{equation}\label{eq:unitary-t}
U(t)=\begin{pmatrix}
\cos\beta_t & \sin\beta_t\\
-\sin\beta_t & \cos\beta_t
\end{pmatrix},
\end{equation}
with 
\begin{equation}\label{eq:tilt-t}
  \tan2\beta_t=\frac{t\sin2\alpha}{\cos^22\alpha+\gamma\sin^22\alpha}.
\end{equation} 
It follows that every instrument that achieves average success
probability $P_t$ must cause \emph{at least} an average disturbance
\begin{equation}\label{eq:dist1-t}
  D_t
=\frac 12\left(1-t\sin2\alpha\sin2\beta_t\right)
+\frac{\cos2\beta_t}4\left[\gamma(\cos4\alpha-1)-\cos4\alpha-1\right].
\end{equation}

Just by varying the control parameter $t$, it is possible to smoothly
move between the limiting cases. For $t=0$, we obtain the identity
map, that is, the no-measurement case. For $t=1$, we obtain Helstrom's
instrument in Eq.~(\ref{eq:Ui}), and 
Eq.~(\ref{eq:tilt-t}) reproduces the tilt given in Eq.~(\ref{eq:tilt}).
However, the crucial difference between Helstrom's limit ($t=1$) and
the intermediate cases is that, for $t<1$, the optimal instrument
\emph{cannot} be interpreted by means of a measure-and-prepare scheme,
and the unitaries $U(t)$ and $U^\dag (t)$ in Eq. (\ref{eq:kraus-t})
represent feedback rotations for outcomes $1$ and $2$.

By eliminating the parameter $t$ from Eqs.~(\ref{eq:prob1-t})
and~(\ref{eq:dist1-t}), one can obtain the optimal tradeoff between
information and disturbance, for any value of $\alpha $
\cite{fuco,qph}.

\section{Experimental schemes for the minimum-disturbing measurement}

In this section we want to show two experimental schemes 
for the realization of the minimum-disturbing measurement. 
The two-level input
system is
encoded on photons degrees of freedom. Since we are interested not
only in the success probability but also in the posterior state of the
system \emph{after} the measurement, we have to focus on indirect
measurement schemes, in which the system is previously made interact
with a probe, and, after such interaction, a projective measurement is
performed on the probe. The \emph{mathematical} parameter $t$
controlling the tradeoff in Eq.~(\ref{eq:prob1-t}) can then be put in
correspondence with a \emph{physical} parameter controlling the
strength of the interaction between the system and the probe. The case
$t=0$ means that the interaction is actually factorized and that the
subsequent measurement on the probe does not provide any information
about the system and the latter is completely unaffected by the
probe's measurement. This is precisely the no-measurement case. On the
contrary, $t=1$ identifies a \emph{completely entangling} interaction,
or, in other words, a situation in which a measurement on the probe
gives the largest amount of information about the system, consequently
causing the largest disturbance. In the following, two possible
settings are discussed: the first one, which is deterministic and
involves the \emph{dual-rail} representation of qubits~\cite{nielsen},
and the second one which is probabilistic and involves the qubit
encoding on the polarization state of a single photon.

\subsection{Deterministic scheme ($t\ll 1$)}

In Ref.~\cite{nielsen} it is shown how to achieve a
maximally entangling gate of the C-NOT type, i.~e.
\begin{equation}
  |i\>_s|1\>_p\mapsto|i\>_s|i\>_p,\qquad i=1,2
\end{equation}
by combining, in the dual-rail representation of qubits, two Hadamard
gates with a non-linear interaction caused by a Kerr medium coupling
the two modes $s$ (system) and $p$ (probe). More explicitly, by
varying the interaction time (or length) between the system mode and
the probe mode inside the Kerr medium, it is possible to achieve the
following unitary evolution:
\begin{equation}
U(\phi)=\begin{pmatrix}
1 & 0 & 0 & 0\\
0 & 1 & 0 & 0\\
0 & 0 & \frac{1+e^{i\phi}}2 & \frac{1-e^{i\phi}}2\\
0 & 0 & \frac{1-e^{i\phi}}2 & \frac{1+e^{i\phi}}2
\end{pmatrix}.
\end{equation}
The two limiting cases correspond to $\phi=0$ for which
$U(0)=\openone$, and $\phi=\pi$ for which $U(\pi)$ realizes a perfect
C-NOT gate. Then, to measure the von Neumann observable
$\{|1\>\<1|,|2\>\<2|\}$ on the probe is equivalent to apply the
instrument in Eq.~(\ref{eq:kraus-t}) onto the system, with $t=\sin^2
(\phi/2)$. The feedback unitary rotation~(\ref{eq:unitary-t}) can be
subsequently applied conditional to the probe measurement outcome.

This scheme is \emph{deterministic}, that is, no events have to be
discarded. However, approaching the limiting value $\phi=\pi$ (or,
equivalently, $t=1$) is quite hard, since too large nonlinearity is needed 
~\cite{ima}. Hence, such a setup can be useful only for
regimes with $t\ll 1$.

\subsection{Probabilistic scheme}

The second proposal is a modification of the setup already
used in Ref.~\cite{dema} to experimentally realize a \emph{universal}
minimum-disturbing measurement. With respect to Ref.~\cite{dema}, only
the feedback rotations are different. This setup has the great
advantage of being completely achievable by linear optics. 
In order to entangle the system with the probe, it needs an
\emph{entangling measurement} to be performed on the joint
system-probe--state. Such a measurement is in fact a parity check,
namely  
a measurement of the observable
\begin{equation}
  \{P_y=|1\>\<1|_s\otimes|1\>\<1|_p+|2\>\<2|_s\otimes|2\>\<2|_p,\ \ 
  P_n=\openone-P _y \}.  
\end{equation}
However, since the outcome ``$n$'' corresponds to a situation in which
the input photon and the probe photon are indistinguishable, we are
forced to post-select just one half of the events, discarding those
corresponding to the outcome $n$. A part of this major drawback,
limiting the actual usefulness of such a measuring instrument in
practical application, using this setup it is possible to explore the
whole range of the parameter values $t\in[0,1]$, contrarily to what
happens using non-linear media.

\section*{Acknowledgments}
F.~B. acknowledges Japan Science and Technology Agency for support
though the ERATO-SORST Project on Quantum Computation and Information.
M.~F.~S. acknowledges MIUR for partial support through PRIN 2005.

\end{document}